\documentclass{article}
\usepackage{graphicx}
\begin{document}
\title{Pulsar timing arrays within rotating and expanding Universe}
\author{Davor Palle \\
ul. Ljudevita Gaja 35, 10000 Zagreb, Croatia \\
email: davor.palle@gmail.com}
\maketitle
\begin{abstract}
{Recent measurements of the four pulsar timing arrays were 
interpreted as a signal of the low frequency stochastic gravitational
wave background. We show that the amplitude and angular correlations 
of pulsar timing residuals can be interpreted as a consequence of the
vortical geodetic motions of pulsar's photons within rotating and expanding
Universe. The resulting angular correlation curves are similar to the Hellings-Downs
curve and the observed amplitude allows the estimate of the vorticity
of the Universe. We show that the estimated vorticity is compatible with the observed
rotation of the CMB polarization vector.}
\end{abstract}

\section{Introduction and motivation}
We are witnessing the recent announcement of the positive signals at four 
pulsar timing arrays (PTA) \cite{PTAs} that are interpreted as a signal of
the stochastic gravitational wave background. However, there are strong
theoretical arguments against the existence of the gravitational waves
in the general theory of relativity (GR) or in the Einstein-Cartan (EC)
theory of gravity \cite{Palle1}. The wave equations exist only in the
linearization procedures when the invariance on the general coordinate
transformations is lost and consequently any strict mathematical relationship to the GR or
EC gravity.

The confusion is now even worse because the LIGO Collaboration announced 
detection of gravitational waves without pointing out to any  
potential sources that can be located via electromagnetic, neutrino 
or cosmic rays. The alternative interpretation of the LIGO events as
the geophysical phenomenon is the action of the ocean tidal bulges 
on the LIGO antenna \cite{Palle2}.

In this paper we show that the signals from PTAs can be described as
an inevitable consequence of the genuine cosmic phenomena: cosmic expansion,
cosmic rotation (vorticity) and cosmic acceleration (not to be confused 
with the "acceleration" of the $\Lambda CDM$ cosmology).
 
The next section is devoted to the formalism of the shearless spacetime
with expansion, vorticity and acceleration. The solution of the geodetic
equation with the support of the Killing vectors is presented.
The formulas for the pulsar's timing residuals and their angular 
correlations are given.
We present and discuss our results in the concluding section.

\section{Geodetic motions in a rotating Universe}
The cosmic expansion is not a phenomenon that should be studied only on
the cosmic scales. For example, even the lunar recession is affected by
the Friedmann-Lema\^{i}tre-Hubble expansion law \cite{Maeder}.

The cosmic acceleration has its imprint in the solar system \cite{Palle3} or
in the CMB \cite{Palle4} as a small tilt of the spectrum. However, it is also
observed in dynamics of wide binary stars \cite{Xavier,Chae} and globular cluster
systems \cite{Bilek} at the galactic scale.

Similarly, the cosmic rotation (vorticity) implies a violation of the
isotropy observed by WMAP \cite{WMAP} and PLANCK \cite{PLANCK} on CMB TT spectrum,
as well as on the CMB polarization \cite{Minami}. The anisotropy is found
in galaxy spin directions \cite{Shamir} and in the anomalous large-scale
flows \cite{flow}.

On the theoretical side we refer the reader to the Einstein-Cartan cosmology
\cite{Palle5} as a framework to resolve cosmological problems without inflaton
scalar introducing rotational degrees of freedom through spin and torsion.
The primordial Majorana light neutrinos are responsible for the appearance
of right-handed vorticity of the Universe \cite{Palle5}.

Let us define the shearless GR geometry with expansion, vorticity and acceleration
\cite{Korotky}:

\begin{eqnarray}
d s^{2}=d t^{2}-R^{2}(t)[d x^{2}+(1-\lambda^{2})e^{2mx}d y^{2}
+d z^{2}]-2 R(t)\lambda e^{mx}d y dt, \\
m,\lambda=const., \nonumber
\end{eqnarray}
\begin{eqnarray*}
g_{\mu\nu} = v_{\mu}^{a}v_{\nu}^{b}\eta_{a b},\ 
\eta_{a b}=diag(+1,-1,-1,-1),\\  
 \mu,\nu=0,1,2,3, \ a,b=\hat{0},\hat{1},\hat{2},\hat{3},\\
 v_{0}^{\hat{0}}=1,\ v_{0}^{\hat{2}}=-\lambda e^{m x}R(t),
 \ v_{1}^{\hat{1}}=v_{3}^{\hat{3}}=R(t),\ v_{2}^{\hat{2}}=e^{m x}R(t).
\end{eqnarray*}

According to Ehlers decomposition, one obtains \cite{Palle5}:

\begin{eqnarray}
\nabla_{\mu}u_{\nu}=\omega_{\nu\mu}
+\sigma_{\mu\nu}+\frac{1}{3}\Theta h_{\mu\nu}+u_{\mu}a_{\nu}, \\
u^{\mu}u_{\mu}=1,\ h_{\mu\nu}=g_{\mu\nu}-u_{\mu}u_{\nu},
\ a_{\mu}=u^{\nu}\nabla_{\nu}u_{\mu}, 
\ \Theta=\nabla_{\nu}u^{\nu}, \nonumber \\
\omega_{\mu\nu}=h^{\alpha}_{\mu}h^{\beta}_{\nu}
\nabla_{[\beta}u_{\alpha ]},\ 
\sigma_{\mu\nu}=h^{\alpha}_{\mu}h^{\beta}_{\nu}
\nabla_{(\alpha}u_{\beta )}-\frac{1}{3}\Theta h_{\mu\nu},
\nonumber 
\end{eqnarray}
\begin{eqnarray*}
  [\alpha\beta]\equiv\frac{1}{2}(\alpha\beta-\beta\alpha),\ 
(\alpha\beta)\equiv\frac{1}{2}(\alpha\beta+\beta\alpha). \nonumber
\end{eqnarray*}

Using the assumed metric of Eq.(1), the expansion, vorticity and 
acceleration take the form (c=velocity of light):

\begin{eqnarray}
H\equiv\frac{\dot{R}}{R},\ \omega\equiv(\frac{1}{2}\omega_{\mu\nu}\omega^{\mu\nu})^{1/2}=
\frac{m \lambda}{2 R}c,\ a\equiv(-a_{\mu}a^{\mu})^{1/2}=\lambda\frac{\dot{R}}{R}c.
\end{eqnarray}

$\lambda$,m,x,y and z are dimensionless physical quantities.

The geodetic equation in GR for a massless photon 
appears to be:

\begin{eqnarray}
\frac{d G^{\mu}}{d \tau}+\Gamma^{\mu}_{\nu\lambda}G^{\nu}G^{\lambda}=0,  \\
G^{\mu}=\frac{d x^{\mu}}{d \tau},\ G^{\mu}G_{\mu}=0. \nonumber
\end{eqnarray}

The presumed geometry implicates three Killing vectors \cite{Korotky}:
\begin{eqnarray}
\xi^{\mu}_{(1)}=(0,\frac{1}{m},-y,0),\ \xi^{\mu}_{(2)}=(0,0,1,0),\ 
\xi^{\mu}_{(3)}=(0,0,0,1).
\end{eqnarray}

The Killing and geodetic equations imply conservation equations:
\begin{eqnarray}
G^{\mu}\nabla_{\mu}(G^{\nu}\xi_{(i)\nu})=0,\ i=1,2,3.
\end{eqnarray}

Defining the conserved quantities as $q_{i}=-\xi^{\mu}_{(i)}G_{\mu},\ i=1,2,3$,
it follows \cite{Korotky}:
\begin{eqnarray}
G^{1}&=&\frac{m}{R^{2}}(q_{1}+y q_{2}), \nonumber \\
G^{2}&=&\frac{1}{(1-\lambda^{2}e^{m x}R^{2}}(\frac{q_{2}}{e^{m x}}-\lambda R G^{0}),
\nonumber \\
G^{3}&=&\frac{q_{3}}{R^{2}}, \nonumber \\
(RG^{0})^{2}&=&(1-\lambda^{2})[m^{2}(q_{1}+y q_{2})^{2}
+\frac{1}{(1-\lambda^{2})e^{2 m x}}q_{2}^{2}+q_{3}^{2}],
\end{eqnarray}

where the last equation is the lightlike condition $G^{\mu}G_{\mu}=0$.
After change of variables, from the affine parameter $\tau$ to $t=x^{0}$ variable and 
$G^{0}=\frac{d x^{0}}{d \tau}$, we finally reach a coupled system of nonlinear
differential equations:
\begin{eqnarray}
R\frac{d x}{d t}&=&\frac{m}{R G^{0}}(q_{1}+y q_{2}), \nonumber \\
R\frac{d y}{d t}&=&\frac{1}{(1-\lambda^{2})e^{2 m x}}(\frac{q_{2}}{R G^{0}}
-\lambda e^{m x}), \nonumber \\
R\frac{d z}{d t}&=&\frac{q_{3}}{R G^{0}}.
\end{eqnarray}

This nonlinear non-stiff system of equations is solved numerically with the conformal time
variable $d \kappa= d t/R(t)$ ($R G^{0}$ is defined in Eq. (7)):
\begin{eqnarray}
\frac{d x}{d \kappa}&=&\frac{m}{R G^{0}}(q_{1}+y q_{2}), \nonumber \\
\frac{d y}{d \kappa}&=&\frac{1}{(1-\lambda^{2})e^{2 m x}}(\frac{q_{2}}{R G^{0}}
-\lambda e^{m x}), \nonumber \\
\frac{d z}{d \kappa}&=&\frac{q_{3}}{R G^{0}}.
\end{eqnarray}

The null-geodesics are parametrized at the observer's position by the spherical angles
\cite{Korotky}:
\begin{eqnarray}
G^{\hat{0}}&=&1,\ G^{\hat{1}}=\sin \theta \cos \phi,\ 
G^{\hat{2}}=\sin \theta \sin \phi,\ G^{\hat{3}}=\cos \theta \nonumber \\
 \Rightarrow q_{1}&=&\frac{R_{0}}{m}\sin \theta \cos \phi , \nonumber \\
q_{2}&=&R_{0}(\sin \theta \sin \phi + \lambda), \nonumber \\
q_{3}&=&R_{0} \cos \theta .
\end{eqnarray}

The frequency is calculated by the wave vector \cite{Ehlers}:
\begin{eqnarray}
G^{\mu}=\frac{d x^{\mu}}{d \tau}=\frac{d t}{d \tau}\frac{d x^{\mu}}{d t}
= G^{0} u^{\mu}, \nonumber \\
u^{\mu}u_{\mu}=1\ \Rightarrow \nu=G^{\mu}u_{\mu}=G^{0},
\end{eqnarray}

thus the redshift turns out to be:
\begin{eqnarray}
\frac{\delta \nu}{\nu_{0}}=\frac{G^{0}(\kappa_{0},x=y=z=0)-G^{0}(\kappa,x,y,z)}
{G^{0}(\kappa_{0},x=y=z=0)}.
\end{eqnarray}

Now we define the anomalous residual in the pulse arrival time:
\begin{eqnarray}
R(T) = \int^{T}_{0} d t [\nu_{0}-\nu (t)]/\nu_{0}.
\end{eqnarray}

Since the metric in Eq.(1) describes a rotation around z-axis and the pulsar 
coordinates are $x_{P}=\sin \theta_{P} \cos \phi_{P}$, $y_{P}=\sin \theta_{P} \sin \phi_{P}$,
$z_{P}=\cos \theta_{P}$, we want to allow the possibility that the array of pulsars 
rotates around arbitrary positioned axis of rotation in the observer's frame
$(\theta_{A},\phi_{A})$.

Let us define the unitary transformation U:

\begin{eqnarray}
U = \left( \begin{array}{ccc}
    -\sin \phi_{A} & -\cos \theta_{A}\cos \phi_{A} & \sin \theta_{A} \cos \phi_{A} \\
    \cos \phi_{A} & -\cos \theta_{A} \sin \phi_{A} & \sin \theta_{A} \sin \phi_{A} \\
    0 & \sin \theta_{A} & \cos \theta_{A}
           \end{array} \right).
\end{eqnarray}

Evidently, one gets:
\begin{eqnarray}
U \left( \begin{array}{c}
  0 \\ 0 \\ 1 \end{array} \right) = 
  \left( \begin{array}{c}
  \sin \theta_{A} \cos \phi_{A} \\ \sin \theta_{A} \sin \phi_{A} \\ \cos \theta_{A} \end{array} \right).
\end{eqnarray}

If we want to use our differential equations for arbitrary axis
of rotation of the Universe, we have to transform the coordinates
of pulsars according to:

\begin{eqnarray}
\left( \begin{array}{c}
x'_{P} \\ y'_{P} \\ z'_{P} \end{array} \right)=
U^{-1} \left( \begin{array}{c}
  x_{P} \\ y_{P} \\ z_{P} \end{array} \right) ,\ \  U^{-1}=U^{\dagger}=U^{T}.
\end{eqnarray}

The relative positions between axis of rotation and pulsars are preserved
and we use the above differential equations with transformed coordinates 
of pulsars $(\theta'_{P},\phi'_{P})$.

\section{Anomalous timing residuals, angular correlations and conclusions}
The set of cosmological parameters used in calculations consists of
(subscript 0 denotes present values):
\begin{eqnarray*}
H_{0}=100 h\ km s^{-1} Mpc^{-1},\ h=0.74, \\
R_{0} = c\ t_{U} \simeq c\ H_{0}^{-1},\ t_{U}\simeq 13.9 Gyr.
\end{eqnarray*}

The time integrations are performed within our galaxy (thus R(t)
is constant to a very good approximation) and within small time intervals:
\begin{eqnarray*}
\kappa-\kappa_{0}=c \int^{t}_{t_{0}} \frac{d t'}{R(t')} = \frac{c}{R_{0}}(t-t_{0}).
\end{eqnarray*}

The residuals vanish
together with vanishing $q_{1}$ or $q_{2}$ and they are proportional
to the square of integrated time $R(T) \propto T^{2}$.

It is important to stress that the anomalous residuals are linearly
proportional to the vorticity $R(T) \propto \omega_{0}$.
One can easily resolve geodesics with nonvanishing acceleration
$a_{0}\ne 0$ and vanishing vorticity $\omega_{0}=0$ and verify that 
residuals vanish. Thus, we infer that the nonvanishing anomalous timing
residuals are genuine consequence of the vortical motions of photons of 
the cosmological origin.

We define N=100 pulsar coordinates isotropically distributed on the 
sphere $\theta_{i} \epsilon [0,\pi),\ \phi_{j} \epsilon [0,2 \pi),\ 
i,j=1,...,10$. This pulsar array serves to evaluate $\frac{N (N-1)}{2}=4950$
angular correlations of each pair of pulsars. The results are 
depicted in Fig.1.

\begin{figure}[htb]
\centerline{
\includegraphics[width=12cm]{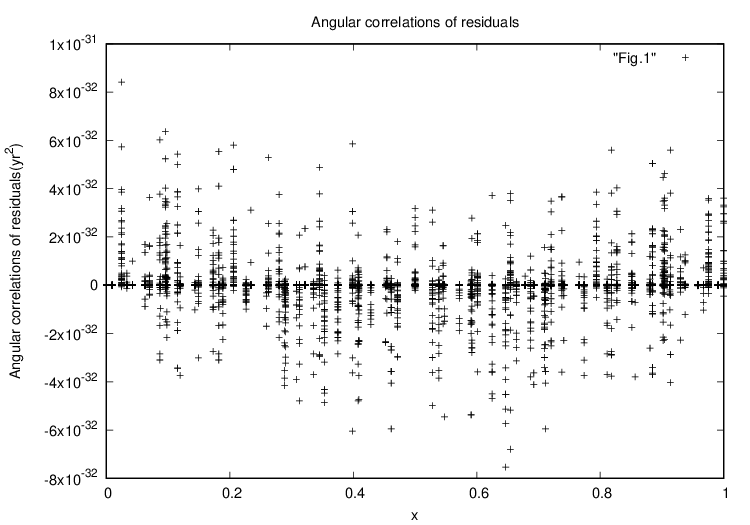}}
\caption{Angular correlations of timing residuals: $\theta_{A}=\phi_{A}=0,\ 
m=10^{-4},\ \lambda=10^{-1},\ T=1yr.$}
\end{figure}
 
We do not average over the axis of rotation since the violation of isotropy
is the essential property of the cosmological model. However, we perform
the Marquardt's method polynomial fit of data with three parameters (see Fig.2,
$a_{0}+a_{1}x+a_{2}x^{2},\
x=\frac{1}{2}(1-\cos \theta_{ij}),\ \cos \theta_{ij}=
\hat{r}_{P,i}\cdot\hat{r}_{P,j})$:

\begin{eqnarray}
<R(T)_{i}R(T)_{j}>_{fit}=\sigma^{2}_{rot}(T)\eta (x),\ \eta(0)=\frac{1}{2}.
\end{eqnarray}

The reader can notice the similarity of the curve $\eta (x)$ with
the Hellings-Downs curve $\zeta (x)$.

\begin{figure}[htb]
\centerline{
\includegraphics[width=12cm]{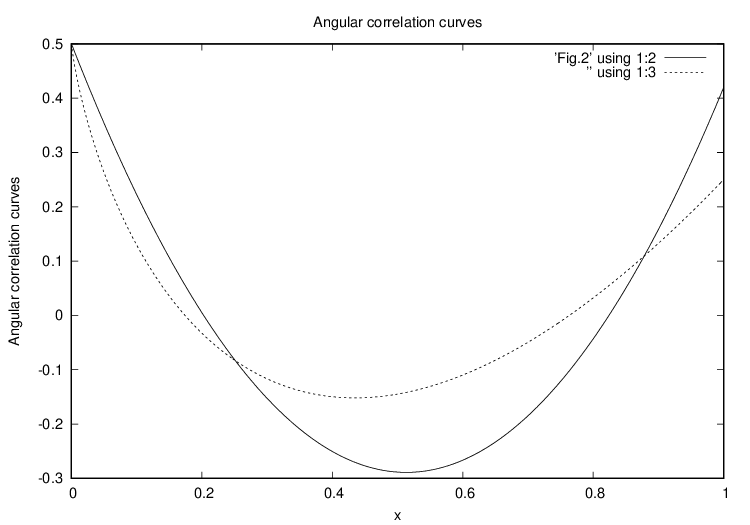}}
\caption{Solid curve is $\eta (x)$ and dashed line is the Hellings-Downs
$\zeta (x)$ curve. Parameters for $\eta$ are the same as in Fig.1 and $\sigma^{2}_{rot}=1.27\times 10^{-32} yr^{2}$.}
\end{figure}

The $\eta (x)$ curve of the residual correlations when the axis of rotation
is not at the z-axis, but with the isotropic array of pulsars as in Fig.1, is
depicted in Fig.3 with dashed line Hellings-Downs curve for comparison.
The results deviate from Fig.2 because of a small number of pulsars in the array when the angular
distances between the pulsars are too large to represent the real statistical isotropic 
set.

\begin{figure}[htb]
\centerline{
\includegraphics[width=12cm]{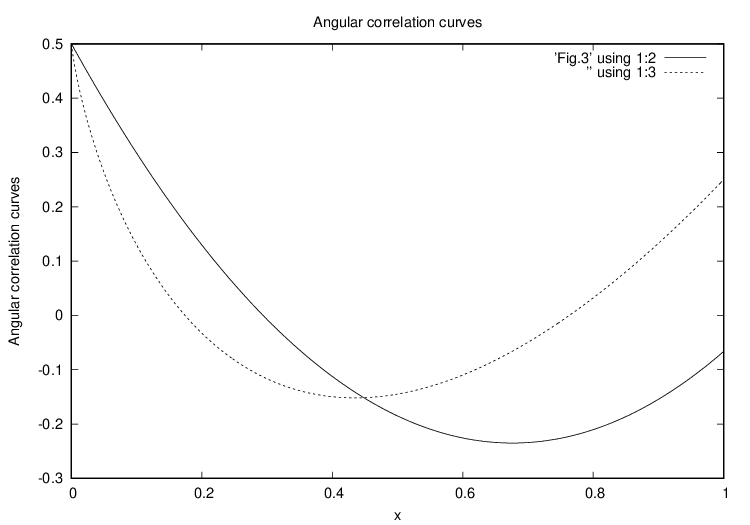}}
\caption{Solid curve is $\eta (x)$ with dashed line Hellings-Downs
$\zeta (x)$ curve. Parameters are $\theta_{A}=0.7,\ \phi_{A}=-2.4,\
m=10^{-4},\ \lambda=10^{-1},\ T=1yr$ and
$\sigma^{2}_{rot}=1.585 \times10^{-32} yr^{2}$.}
\end{figure}

If the pulsars are not isotropically distributed across the sky, the correlations
receive complicated distributions.

The four pulsar timing arrays \cite{PTAs} interpret their data 
by the well established calculations of the action of gravitational
waves on pulsar's photons \cite{Sazhin,Detweiler,Hellings}:
\begin{eqnarray}
<R(T)_{i}R(T)_{j}>_{average}=\sigma^{2}_{gw}(T)\zeta (x),\ \zeta(0)=\frac{1}{2}, \\ 
\zeta(x)=\frac{1}{2}-\frac{x}{4}+\frac{3}{2} x \ln x = Hellings-Downs\ curve, 
\nonumber \\
\sigma^{2}_{gw}(T)=\int^{f_{h}}_{f_{l}} d f \frac{1}{12 \pi^{2}}\frac{A^{2}}{f^{3}}
(\frac{f}{1 yr^{-1}})^{2 \alpha}
\simeq \frac{A^{2}}{12 \pi^{2}}\frac{1}{2-2\alpha}(f_{l}/1 yr^{-1})^{2\alpha-2}
yr^{2}, \\
f_{l}=\frac{1}{T(yr)},\ f_{l} \ll f_{h} \Rightarrow \sigma^{2}_{gw}(T)
\propto T^{2-2 \alpha}. \nonumber
\end{eqnarray}

From the fact that $<R(T)_{i}R(T)_{j}>_{fit} \simeq <R(T)_{i}R(T)_{j}>_{average}$
and the average amplitude of four PTAs \cite{PTAs},
we estimate the magnitude of the vorticity of the Universe in GR:
\begin{eqnarray}
x=0,\ frequency=1 yr^{-1},\ A=2.24\times 10^{-15},\ \alpha=-\frac{2}{3} 
\nonumber \\
\Rightarrow \omega_{0}={\cal O} (10^{-5} H_{0}).\ \ \ \ \ 
\end{eqnarray}

At the end we draw few conclusions: \\
(1) the first positive signal from the PTAs \cite{PTAs} can be elucidated as
a consequence of the vortical motions of pulsar's photons in a rotating
Universe without any reference to the stochastic gravitationl wave background, \\
(2) timing residuals are proportional to the total time span $R_{rot}(T) \propto T^{2}$
and to the vorticity $R_{rot}(\omega_{0}) \propto \omega_{0}$, \\
(3) two parameters of the metric in Eq.(1) m,$\lambda$ allow to fix two
observables $\omega_{0}$ and $a_{0}$ (see Eq.(3)); for example if 
$m=10^{-4}$ and $\lambda=10^{-1}$, one gets $\omega_{0}={\cal O}(10^{-5} H_{0})$
and $a_{0}={\cal O}(10^{-10} m s^{-2})$ \cite{Palle3,Palle4,Chae,Bilek}, \\
(4) although our considerations are in GR, a definition of the effective
vorticity in the EC cosmology \cite{Palle5},
\begin{eqnarray*}
\tilde{\omega}_{\mu\nu}=h^{\alpha}_{\mu}h^{\beta}_{\nu}
\tilde{\nabla}_{[\beta}u_{\alpha ]},
\end{eqnarray*}
causes a relation $|\tilde{\omega}_{0}|=|\omega_{0}-Q_{0}|={\cal O} (10^{-5}H_{0})$,
where $Q^{2}=\frac{1}{2}Q_{\mu\nu}Q^{\mu\nu}$ and $Q^{\alpha}_{\cdot \mu\nu}
=u^{\alpha}Q_{\mu\nu}=torsion$. This is in accord with the 
geodesic equation in the EC cosmology:
\begin{eqnarray*}
\frac{d G^{\mu}}{d \tau}+\tilde{\Gamma}^{\mu}_{(\nu\lambda)}G^{\nu}G^{\lambda}=0.
\end{eqnarray*}
However, one should accept a very strong argument \cite{Hehl} that the photon as a massless
particle cannot be coupled to the torsion, hence $|\tilde{\omega}_{0}|=|\omega_{0}|$, \\
(5) knowing the vorticity of the Universe we can estimate the rotation $\beta$ of the CMB polarization
vector integrating the infinitesimal rotations $\delta \beta = -\Im \rho \delta s$ \cite{Korotky},
where $\rho = spin\ coefficient,\ s= affine\ parameter$, from the decoupling to the present time
($\Im \rho=\omega,\ s=R_{0}a$ and \cite{Peebles,Palle6}: $\omega(a)\propto a^{-2}$):
\begin{eqnarray}
\beta=-R_{0}\omega_{0}\int^{1/(1+z_{dec})}_{1}a^{-2}da
=R_{0}\omega_{0}z_{dec}.
\end{eqnarray}
Because $R_{0}\simeq H_{0}^{-1},\ z_{dec}\simeq 1100$ and Eq.(20), we have $\beta={\cal O}(10^{-2})rad$
that is compatible with measurements $\beta=\simeq 0.215^{\circ}\ to\  \simeq 0.35^{\circ}$ \cite{Minami}.
The positivity of the CMB polarization rotation angle confirms the prediction of
the Einstein-Cartan cosmology that the rotation of the Universe has right handed chirality \cite{Palle5}.

\end{document}